\documentclass[aps,prc,floats,floatfix,twocolumn,superscriptaddress,nofootinbib]{revtex4}

\usepackage{epsfig}
\newcommand{\nuc}[2]{$^{#1}${#2}}
\newcommand{\etal}{\emph{et al.}}
\newcommand{\nn}{\nonumber}

\def\be{\begin{equation}}
\def\ee{\end{equation}}
\def\rb{\rangle}
\def\lb{\langle}
\def\twoplus{{2$^+$}~}
\def\transition{{$ 2^+ \to 0^+$}~}
\def\rme{\lb 0 || \hat{Q}_2 || 2 \rb}
\def\numA{359~}
\begin{document}

\title{Global study of the spectroscopic properties
       of the first $2^+$ state in even-even nuclei}

\author{B. Sabbey}
\affiliation{Department of Physics and Institute for Nuclear Theory,
             Box 351560, University of Washington, Seattle, WA 98195}

\author{M. Bender}
\affiliation{Dapnia/SPhN, CEA Saclay,
             F-91191 Gif sur Yvette Cedex,
             France}
\affiliation{Centre d'Etudes Nucl{\'e}aires de Bordeaux Gradignan,
             BP120, F-33175 Gradignan Cedex, France}

\author{G. F. Bertsch}
\affiliation{Department of Physics and Institute for Nuclear Theory,
             Box 351560, University of Washington, Seattle, WA 98195}

\author{P.-H. Heenen}
\affiliation{Service de Physique Nucl{\'e}aire Th{\'e}orique,
             Universit{\'e} Libre de Bruxelles,
             CP 229, B-1050 Brussels, Belgium}

\date{November 22 2006}


\begin{abstract}
We discuss the systematics of the $2^+$ excitation energy and the
transition probability from this $2^+$ to the ground state  for
most of the even-even nuclei, from \nuc{16}{O} up to the
actinides, for which data are available.
To that aim we calculate their correlated \mbox{$J=0$} ground
state and \mbox{$J=2$} first excited state by means of the
angular-momentum and particle-number projected generator
coordinate method, using the axial mass quadrupole moment as the
generator coordinate and self-consistent mean-field states only
restricted by axial, parity, and time-reversal symmetries.
The calculation, which is an extension of a previous systematic
calculation of correlations in the ground state, is performed
within the framework of a non-relativistic self-consistent
mean-field model using the same Skyrme interaction SLy4 and a
density-dependent pairing force to generate the mean-field
configurations and mix them.
To separate the effects due to angular-momentum projection and
those related to configuration mixing, the comparison with the
experimental data is performed for the full calculation and also
by taking a single configuration for each angular momentum, chosen
to minimize the projected energy. The theoretical energies span
more than 2 orders of magnitude, ranging below 100 keV in deformed
actinide nuclei to a few MeV in doubly-magic nuclei. Both
approaches systematically overpredict the experiment excitation
energy, by an average factor of about 1.5. The dispersion around
the average is significantly better in the configuration mixing
approach compared to the single-configuration results, showing the
improvement brought by the inclusion of a dispersion on the
quadrupole moment in the collective wave function. Both methods do
much better for the quadrupole properties; using the configuration
mixing approach the mean error on the experimental $B(E2)$ values
is only 50$\%$. We discuss possible improvements of the theory
that could be made by introducing other constraining fields.
\end{abstract}

\maketitle
%
%
\section{Introduction}
\label{sect:intro}

Self-consistent mean-field methods (SCMF) are the only
computationally tractable methods which can be applied to medium and
heavy nuclei and have a well-justified foundation in many-body
theory~\cite{RMP}. Recently there has been considerable progresses
in using these methods to compute nuclear mass tables~\cite{doba}.
One of the appealing features of the SCMF is that the properties of
all nuclei are derived from a fixed energy density functional that
depends on a small number of universal parameters, and can be used
for the entire chart of nuclei.  The Skyrme family of functionals
which is used in the present study, depends on about 10 parameters
for the particle-hole interaction with 2-3 extra parameters for the
pairing interaction.

In a previous study~\cite{Ben05a,Ben06}, we have used an extended
SCMF theory to calculate the ground state binding energies of the
about $600$ even-even nuclei whose masses are known experimentally.
In particular, two extensions were introduced to treat correlation
effects going beyond a mean field approach. The first is a
projection of the SCMF wave functions to restore symmetries broken
by the mean field: particle numbers and angular momentum. The second
is a mixing of projected mean-field states corresponding to
different intrinsic axial quadrupole deformations. These
calculations were performed with the same energy functional as for
the determination of the mean-field configurations, so they do not
require to introduce new parameters. Our main aim was to determine
the effect of correlations on masses. In particular, the error on
experimental masses in microscopically based methods presents arches
between the magic numbers. The correlations added in
Refs.~\cite{Ben05a,Ben06} clearly reduce the amplitude of the arches
in the mass residuals, but do not remove them completely. For the
parameterization we have used, however, the arches are much more
pronounced along isotopic chains than along isotonic chains, which
suggests that their appearance is not only related to missing
correlations, but also to deficiencies of the currently used
effective interactions. There is a clear improvement when looking at
mass differences between neighboring nuclei around magic ones, in
particular when crossing proton shells. Similarly, the systematics
of charge radii is also improved, particularly in the transitional
region between spherical and well-deformed nuclei. Altogether, this
study clearly confirms the importance of incorporating some beyond
mean-field correlations explicitly in the method and not
heuristically in the energy density functional.

In this work, we extend our previous study to two new observables:
the excitation energy of the first $2^+$ state and the $B(E2)$ value
for the transition between the first $2^+$ and the ground state.

A simple approximation that can be systematically applied is to
start from a set of constrained SCMF configurations corresponding to
different axial quadrupole moments and to project them on angular
momentum. Then, one searches the projected configuration that has
the lowest energy for each angular momentum. We will call this
procedure minimization after projection (MAP). A more sophisticated
procedure requires an additional step: for each $J$-value, the total
energy is further minimized by mixing projected SCMF configurations
corresponding to different deformations.  The mixing of constrained
SCMF configurations is called the Generator Coordinate Method (GCM),
and is based on the solution of the Hill-Wheeler (HW) equation. We
shall label configuration mixing calculations by HW.. Although
numerically demanding, this approach has nowadays been used to study
the detailed structure of low-energy collective excitation spectra
of nuclei up to the actinide region using non-relativistic Skyrme
\cite{O-16,Ben03a,Ben04b,Ben04c,Ben05b} and Gogny
\cite{ro02,Rod02a,Egi04a} interactions as well as relativistic
Lagrangians \cite{Nik06a,Nik06b}. Here, however, as in \cite{Ben05a,Ben06},
we aim at something different: the calculation of a few very
specific properties of the lowest $0^+$ and $2^+$ states for several
hundred nuclei. For this purpose, the numerical procedure necessary
for a detailed study by the GCM is too costly to be applied on such
a large scale with present computational resources. Our bias on
lowest collective $0^+$ and $2^+$ states, however, permits to set-up
a tailor-made numerical scheme that reduces the effort considerably.
For the angular momentum projection, we will generalize the
topological Gaussian overlap approximation~\cite{re78,ha02} (GOA)
used in our previous work. The GCM calculations will also be reduced
in size by truncations of the configuration space.
%
%
\section{Calculational Details}
\label{sect:method}
As mentioned in the introduction, calculations are performed
along the lines of Ref.\ \cite{Ben06}. We will briefly summarize
the most important points and give details only for the necessary
extensions to calculate $2^+$ states and matrix elements of the
quadrupole operator. The calculations reported here go beyond mean
field in three respects: (i) projections on good particle numbers;
(ii) projection on angular momentum \mbox{$J=0$} and 2; and (iii)
mixing of states with different intrinsic deformation.
All the results presented in this paper include
particle-number projection and we drop explicit reference to
particle number from the notation.

\subsection{DFT Calculations}
We use the code {\tt ev8} (see Refs.\ \cite{Bon85a,Bon05a}) to solve
the SCMF equations for an energy functional based on the Skyrme
interaction. The single-particle orbitals are discretized on a
three-dimensional Lagrange mesh corresponding to a cubic box in
coordinate space. The code imposes time-reversal symmetry on the
many-body state, assuming pairs of conjugated states linked by
time-reversal and having the same occupation number, which limits
the description to even-even nuclei, and non-rotating states. The
only other restriction on the wave function is that the Slater
determinant of the orbits is invariant with respect to parity and
axial rotations. For this work we take the SLy4 Skyrme
parameterization~\cite{Cha98}, the same as we used in our previous
global study. For the pairing interaction we choose an energy
functional that corresponds to a density-dependent zero-range
pairing force, with cutoffs at 5 MeV above and below the Fermi
energy, as described in \cite{Rig99}. As in earlier projected GCM
studies, the pairing strength is taken to be $-1000$ MeV fm$^3$ for
both protons and neutrons.

To avoid a breakdown of pairing correlations for small level
densities around the Fermi surface, we enforce the presence of
pairing correlations using the Lipkin-Nogami (LN) prescription as
described in Ref.\ \cite{Gal94a}. However, we emphasize that the LN
prescription is only used to generate wave functions of the BCS
form; physical properties are calculated with the code {\tt
promesse}~\cite{promesse}, which performs projections on proton and
neutron particle numbers and provides the matrix elements needed for
angular momentum projection.

Mean-field states with different mass quadrupole moments are generated by
adding a constraint to the mean-field equations to force the
intrinsic axial quadrupole moment $q$  to have a specific value.
Higher-order even axial multipole moments are automatically optimized
for a given mass quadrupole moment.  A typical calculation for a
nucleus involves the construction of about 20 SCMF wave functions
that span a range of deformations sufficient to describe the
ground state.

%
%
\subsection{Beyond mean field}

Formally, eigenstates  $| J M q \rangle$ of the angular momentum operators
$\hat{J}^2$ and $\hat{J}_z$ with eigenvalues $J(J+1)$ and $M$ are obtained
by application of the operator
\begin{equation}
\label{eq:PJ}
\hat{P}^J_{MK}
= \frac{2J+1}{16 \pi^2}
  \! \int_{0}^{4\pi} \! \! d\alpha
  \! \int_0^\pi \! \! d\beta \; \sin(\beta)
  \!  \int_0^{2 \pi} \! \! d\gamma \;
  \mathcal{D}^{*J}_{MK} \,
  \hat{R}
,
\end{equation}
on the SCMF states $| q \rangle$. The rotation operator
$\hat{R}$ and the Wigner function $\mathcal{D}^{J}_{MK}$ both depend
on the Euler angles $\alpha,\beta,\gamma$.

In a further step, we consider the variational configuration
mixing in the framework of the Generator Coordinate Method.
Starting from the \emph{ansatz}
\begin{equation}
| J M k \rangle
= \sum_q f_{Jk} (q) \, | J M q \rangle
,
\end{equation}
for the superposition of projected SCMF states, where $k$ labels the
states for given $J$ and $M$, the variation of the energy
$\langle J M k | \hat{H} | J M k \rangle /
 \langle J M k | J M k \rangle$
leads to the discretized Hill-Wheeler-Griffin equation
\begin{equation}
\label{eq:HWG}
\sum_{q'}
\big[ H_J (q,q') - E_{J,k} \, I_J (q,q') \big] \; f_{J,k} (q')
= 0
\end{equation}
that determines the weights $f_{Jk}(q)$ of the SCMF states in the projected
collective states, and the energy $E_{J,k}$ of the collective states.

We have to calculate diagonal and off-diagonal matrix elements of
the norm and Hamiltonian kernels. For the sake of simple notations,
we use a Hamiltonian operator in all formal expressions, although
there is no Hamiltonian corresponding to an effective energy
functional. In practice, as it is common procedure \cite{Rod02a}, we
replace the local densities and currents entering the mean-field
energy functional with the corresponding transition densities.

The axial symmetry of the mean-field states allows to simplify the
3-dimensional integral over Euler angles to a one-dimensional
integral:
\begin{eqnarray}
\label{eq:norm:kernel:exact}
I_J (q,q')
& = & \langle J M q | J M q' \rangle
      \nn \\
& = & \frac{1}{\mathcal{N}_{Jq} \mathcal{N}_{Jq'}} \!
      \int_0^{1} \! \! d \cos(\beta) \, d^J_{00} (\beta) \,
            \langle q | \hat{R}_\beta | q' \rangle
      \nn \\
      \\
\label{eq:ham:kernel:exact}
H_J (q,q')
& = & \langle J M q | \hat{H} | J M q' \rangle
      \nn \\
& = & \frac{1}{\mathcal{N}_{Jq} \mathcal{N}_{Jq'}} \!
      \int_0^{1} \! \! d \cos (\beta) \, d^J_{00} (\beta) \,
             \langle q | \hat{R}_\beta \hat{H} | q' \rangle
      \nn \\
\end{eqnarray}
with normalization factors
\begin{equation}
\label{eq:norm:proj}
\mathcal{N}_{Jq}^2
= (2 J + 1) \int^1_0 \! d \cos (\beta) \;
  d^J_{00}(\beta) \,
  \langle q | \hat{R}_\beta |  q \rangle
.
\end{equation}
For the calculation of $B(E2)$ values and spectroscopic quadrupole
moments we also have to evaluate matrix elements of the quadrupole
operator, which is outlined in appendix \ref{sect:app:Q2}.
A more detailed discussion of the method can be found
in Refs.~\cite{Ben05b,Ben06} and references given therein.

\subsubsection{topGOA overlaps and Hamiltonian matrix elements}

\begin{figure}
\includegraphics{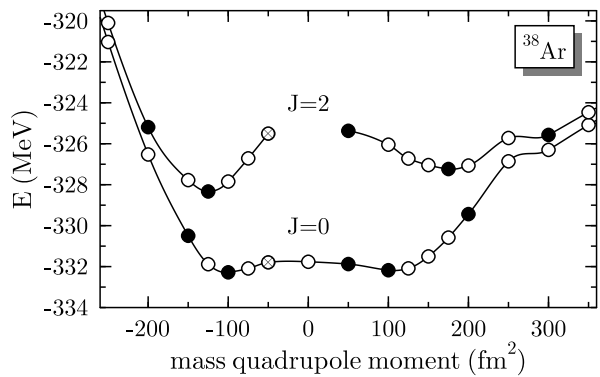}
\caption{\label{Fi:Ar38landscape}
Energy landscapes for $J=0$ and $J=2$ angular-momentum projected
states in $^{38}$Ar.  The open circles show the $q$ values of the
calculated configurations, with lines drawn to guide the eye.
Solid circles are ones included in the mixed configuration calculations
for the global survey.
}
\end{figure}

In Ref.~\cite{Ben04a}, we found that for the description of the
properties of the ground state the $J=0$ projected overlaps can be
computed with a sufficient accuracy for our purpose with a 2- or
3-point approximation to the integral using an extension of the
gaussian overlap approximation called the topGOA \cite{ha03}. There
the rotated overlaps are parameterized by
\begin{equation}
\label{eq:norm:rot}
\lb q | \hat{R}_\beta | q' \rb_t
= \left\{ \begin{array}{l}
         \lb q | q' \rb \, e^{-c(q,q') F(\beta)} \\
         \text{or} \\
         \lb q | q' \rb \, e^{-c(q,q') F(\beta)-d(q,q')F^2(\beta)}
         \end{array}
   \right.
\end{equation}
where $F(\beta) = \sin^2(\beta)$ and the subscript $t$ specifies the
topGOA approximation. This form satisfies the requirement of the GOA
that $F(\beta) \to \beta^2$ for small $\beta$ as well as the
topological requirement that $F(\pi-\beta)=F(\beta)$. Projected
matrix elements of the Hamiltonian are also needed; these were
calculated assuming the functional form
\begin{widetext}
\begin{eqnarray}
\label{eq:goa_H}
\lb q | \hat{R}_\beta \hat{H} | q' \rb_{t2}
& = & \lb q | q' \rb \, e^{-c(q,q') F(\beta)}
      [ h_0(q,q') - h_2 (q,q') F(\beta) ]
      \nn \\
\lb q | \hat{R}_\beta \hat{H} | q' \rb_{t3}
& = & \lb q | q' \rb \, e^{-c(q,q') F(\beta) - d(q,q') F^2(\beta)}
      [ h_0(q,q') - h_2 (q,q') F(\beta) - h_4 (q,q') F^2(\beta) ]
\end{eqnarray}
\end{widetext}
for the 2- and the 3-point approximations respectively. In general,
the 2-point approximation is adequate for heavy nuclei and large
deformations, but the 3-point approximation is necessary to describe
light nuclei. We take points at $\beta$ equal to  zero,  and to a
value where $\lb q | \hat{R}_\beta| q'\rb \approx 0.5$ for the
2-point approximation. A third point is added at $\beta=\pi/2$ for
the 3-point approximation.  This is important for matrix elements
between oblate and prolate configurations, which have their
maximum value at $\pi/2$,

In Fig.~\ref{Fi:Ar38landscape}, we show an example of an energy
curve determined by this procedure for \nuc{38}{Ar}. Angular
momentum projection changes the quadrupole moment from the intrinsic
one to the one observable in the laboratory frame, which now depends
on angular momentum. Most notably, it is zero for $J=0$ states
independent of the deformation of the intrinsic configuration. As a
consequence, it is more convenient and intuitive to use the
intrinsic quadrupole moment of the SCMF states to label the
projected states. The marked points correspond to the $q$ values of
the SCMF configurations that were previously calculated. They are
connected with lines to distinguish the $J=0$ and $J=2$ curves.

The $J=0$ curve  has two very shallow minima at deformations
$q\approx-100$ and $+100$ fm$^2$.  The $J=2$ curve has a pronounced
oblate minimum at $-125$ fm$^2$ and a shallow secondary minimum at
$+175$ fm$^2$. For the MAP calculation, we next estimate the
quadrupole moment at the minimum by interpolating between the
calculated points. We then redo the calculations at the estimated
minimum to find the MAP energy and quadrupole properties. For
\nuc{38}{Ar}, the minimum for $J=0$ is at $q_0= -96$~fm$^2$ with and
energy $E_0= -332.32$~MeV. The corresponding quantities for $J=2$
are $q_2 =-120$~fm$^2$ and $E_2=-328.33$~MeV. The MAP excitation
energy is the difference, $E_{20}\equiv E_2-E_0 = 3.9$~MeV. This is
80 $\%$ higher than the experimental excitation energy of 2.17 MeV.
This is a rather extreme case, in that the $2^+$ of \nuc{38}{Ar} is
very likely better described as a broken-pair two-quasiparticle
state than as a field-induced deformed state. We will return to this
point later.

\subsection{Matrix elements of the quadrupole operator}
\label{sect:Q2:mat}

The calculation of the quadrupole moments of projected states requires
the calculation of all components of the quadrupole tensor.
$Q_{2 \pm1}$ and $Q_{2 \pm2}$ are of course exactly zero for axial
mean-field states with the $z$ axis as symmetry axis as chosen here,
but they have non-vanishing transition matrix elements between
a rotated and an unrotated state.

The detailed expressions for the quadrupole operator and its
projected matrix elements can be found in appendix
\ref{sect:app:Q2}. For axial states, as used here, only the matrix
elements of $\hat{Q}_{20}$ and the real part of the matrix elements
of $\hat{Q}_{21}$ and $\hat{Q}_{22}$ need to be calculated, which
simplifies the computational task.

To calculate matrix elements of the quadrupole operator $\hat{Q}_{2\mu}$, some
modifications of the GOA parameterization are necessary since the
functional behavior of the operator depends on its azimuthal angular
momentum $\mu$. In particular, for the matrix element of $Q_{2\pm 1}$, the form
$F(\beta)=\sin^2(\beta)$ used in the polynomial expansion of
Eq.~(\ref{eq:goa_H}) is not topologically correct.
We therefore
define a topGOA by taking $d^2_{\mu 0}(\beta)$ for the argument of
the polynomial expansion
\begin{eqnarray}
\label{eq:Q2:GOA}
\lefteqn{\Re \big\{ \lb q | \hat{R}_\beta \hat{Q}_{2\mu} | q' \rb \big\}
} \nn\\
& = & \lb q| \hat{R}_\beta | q' \rb_t
      \left[   a_0
             + a_2 d^2_{\mu 0}(\beta)
             + a_4 \left( d^2_{\mu 0}(\beta) \right)^2
      \right]
,
\end{eqnarray}
where the coefficients of the polynomial $a_i$ depend on $q,q'$. As
with the other matrix elements, it is important to include the
point at $\beta=\pi/2$ when $q$ and $q'$ have opposite signs.

There is an additional complication compared to the norm
and Hamiltonian kernels. While for these scalar operators
the kernels (\ref{eq:norm:rot}) and (\ref{eq:goa_H}) are invariant
under exchange of $| q \rangle$ and $| q' \rangle$, this is not
the case for the quadrupole operator, where
$\langle q | \hat{R}_\beta \hat{Q}_{2\mu} | q' \rangle$ is
not equal to $\langle q' | \hat{R}_\beta \hat{Q}_{2\mu} | q \rangle$.
To avoid the explicit calculation of both, we  express the latter
matrix elements as a weighted sum of the former using angular-momentum
algebra and the symmetry properties of the SCMF states \cite{promesse}.
A separate topGOA is then set up to calculate the projected matrix
element from the $\langle q' | \hat{R}_\beta \hat{Q}_{2\mu} | q \rangle$.
It has to be noted, however, that the difference plays a role only
for transition matrix elements between states with different angular
momentum. As we are interested here in $2^+ \to 0^+$ transitions only, the
topGOA for matrix elements with exchanged arguments is needed for $\mu=0$
only, Eq.~(\ref{eq:Q2red:trans}).

\begin{figure}
\includegraphics{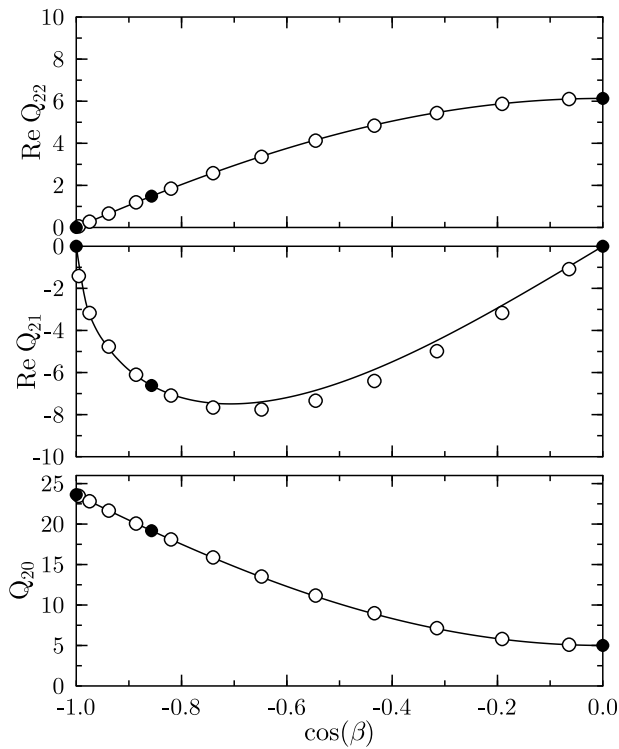}
\caption{\label{fi:cr52q012}
The matrix element $\lb q| \hat R_\beta \hat{Q}_{2\mu} | q \rb$ for the
nucleus \nuc{52}{Cr} at an intrinsic deformation of $q=150$~fm$^2$. The open
circles show the points used for evaluating the integral by a
12-point Gaussian quadrature.  The solid circles show the points
used for the topGOA, and the resulting fit.  The three panels show
the results for $\mu=0,1$ and 2 going up from the bottom panel.
}
\end{figure}

An example for the fits of the integrand is shown in
Fig.~\ref{fi:cr52q012} for \nuc{52}{Cr} at a deformation of $q=q'=
150$~fm$^2$.  Starting from the bottom panel, the three panels show
the rotated overlap matrix element $\lb q| \hat R_\beta
\hat{Q}_{2\mu} | q \rb$ for $\mu=0$, 1 and 2 respectively. The open
circles are the points used to evaluate the integral by a 12-point
Gaussian quadrature, as it would be used in a calculation for the
complete low-energy spectroscopy of this nucleus. The solid circles
are the points used for the topGOA fit, the resulting curves being
indicated by lines. The agreement is excellent; the error associated
with the topGOA is typically less than 1 $\%$ for the $\mu=0$ matrix
element.  This is the only one needed to calculate the $B(E2)$ value
of the \transition transition (see appendix \ref{sect:app:Q2}). The
middle panel shows the integrand for the $Q_{21}$ operator. In
effect, only the middle point can be used for the fit because the
integrand vanishes at $\beta=0$ and $\pi$. Nevertheless, this
approximation works rather well. It is less accurate for some
non-diagonal matrix elements, particularly for matrix elements
connecting configurations with very different deformations which are
needed to describe soft nuclei. The top panel shows the matrix
element for the operator $Q_{22}$. Here there are effectively two
points to determine the topGOA fit.

\begin{figure}
\includegraphics{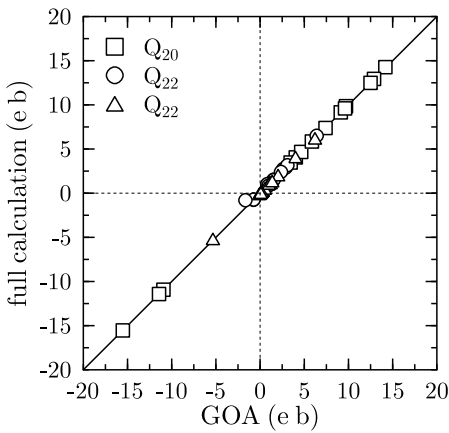}
\caption{\label{Fi:q_goa}
Comparison of the topGOA and full calculation for
$Q_{2\mu}$ for various nuclei and deformations.
}
\end{figure}

To test this approximation further, we have compared the topGOA
quadrupole matrix elements with the matrix elements obtained by a
full integration for a variety of nuclei and deformations.  The
result is shown in Fig.~\ref{Fi:q_goa}.  One can conclude that
Eq.~(\ref{eq:Q2:GOA}) is of sufficient accuracy for our purpose.

\subsection{Configuration mixing}

As mentioned above, we typically compute about $N_q=20$ SCMF
configurations to construct the energy landscape.  For many nuclei,
however, only about half that number can be kept in the
configuration mixing calculation due to ill-conditioned norm
matrices when the space is overcomplete. Nevertheless, the full
configuration mixing calculation requires to compute about 50
projected matrix elements, which is beyond our computational
resources for a study of several hundred nuclei. In
Ref.~\cite{Ben04a}, a GOA was developed for a coordinate
corresponding to the deformation $q$, permitting calculations to be
made to the needed accuracy only using the diagonal and subdiagonal
elements of the configuration mixing matrix, i.e.\ about $2N_q$
projected matrix elements. Unfortunately, the demands on the
approximation are more severe when calculating quadrupole matrix
elements between states of different angular momentum. The matrix
element can change sign, depending on the deformations. For matrix
elements connecting different manifolds of states ($0^+$ and $2^+$),
there is no diagonal element to anchor the GOA.

Lacking a reliable GOA to determine the off-diagonal quadrupole
matrix elements, we took another approach to simplify the
configuration mixing calculation. The number of configurations has
been reduced for each nucleus to a number small enough to make a
global calculation feasible but large enough to have a sufficient
accuracy on the energy of the lowest $0^+$ and $2^+$ states. Since
we have to deal with energy curves of very different topologies,
some care must be taken into the selection of points. The
procedure that we have followed is explained in appendix
\ref{sect:app:HW6}. The number of selected configurations varies
from 3 to 10, but is most often equal to 6. We have therefore
labeled this approximation HW-6.

\begin{figure}[t!]
\includegraphics{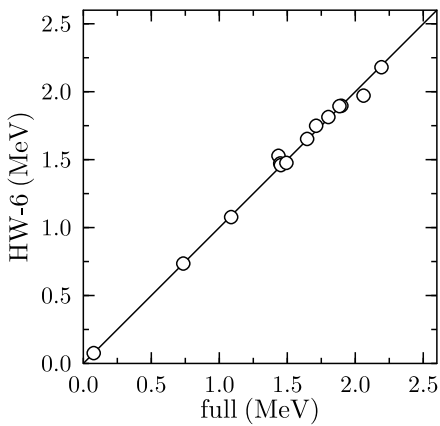}
\caption{\label{fi:6.3}
Comparison of $2^+$ excitation energies for the full
and the HW-6 bases.}
\end{figure}

The points selected  for \nuc{38}{Ar} are shown in
Fig.~\ref{Fi:Ar38landscape} as black circles.  In this case, the
single-configuration MAP energy of the ground state is
$E_0(\text{MAP})=-332.25$~MeV, as quoted in the last section. The gain in
energy from configuration mixing with the large set of
configurations (11
in this case) is $E_0(HW) - E_0(\text{MAP}) = -332.69
+ 332.25 = -0.44$~MeV.  This is to be compared with
$-332.61+332.25= -0.36$ MeV for the HW6 space.
The error, $0.08$ MeV, is within our targeted limit of accuracy.  For
the $J=2$ projected wave functions, the energy gain by the HW
treatment is $-0.68$ MeV and the error of the 6-configuration
truncation is 0.07 MeV with the same sign as in the ground state
energy.
With our present computer resources, we were able to test the HW6
truncation for about 100 nuclei. The $2^+$ excitation energies
computed both ways are compared in Fig.~\ref{fi:6.3}. The
approximation reproduces the energies to an r.m.s.\ accuracy of
better than 0.1 MeV.  The worst cases are two nuclei with coexisting
minima at low excitation energy that are separated by very low
barriers, \nuc{188}{Pb} and \nuc{190}{Pb}, visible as points off the
line at the bottom left-hand corner of the graph.

\begin{figure}[t!]
\includegraphics{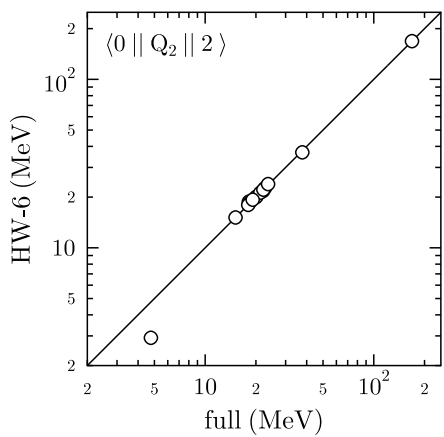}
\caption{\label{fi:6.2}
Comparison of the reduced matrix element
$\langle 0 || \hat{Q}_2 || 2 \rangle$ for the $2^+ \to 0^+$
transition obtained within the full and the HW-6 bases.
}
\end{figure}

The same comparison for the reduced matrix element of the
\transition transition is shown in Fig.~\ref{fi:6.2}.  The agreement
is very good except for the light Pb isotopes, \nuc{182}{Pb},
\nuc{188}{Pb}, and \nuc{190}{Pb}. Among the nuclides, the light Pb
isotopes are rather singular and we shall examine \nuc{188}{Pb} in
more detail in the next section. Overall, the accuracy of the HW-6
approximation is more than adequate for the present global survey.
%
%
\section{Some examples}
\label{sect:benchmarks}

In this section we shall examine the results for a sample of
nuclei with energy maps of different topologies: a light
doubly-magic system, \nuc{40}{Ca}; a heavy doubly-magic system,
\nuc{208}{Pb}; a transitional nucleus near magicity, \nuc{38}{Ar};
a soft nucleus exhibiting triple shape coexistence, \nuc{188}{Pb}; and a
well-deformed heavy nucleus, \nuc{240}{Pu}. We first examine the
results of the MAP approximation.

\subsection{MAP}
The MAP approximation is a variation after projection method within
the very limited subspace defined by the axial quadrupole operator.
For each $J$-value, one finds the configuration leading to the
lowest energy, which thus could be different for $J=0$ and $J=2$.
The results for the observables of interest are presented on the
first line of Table~\ref{ta:examples}, together with the
experimental data on the last line~\cite{ra02}. In all cases, except
\nuc{188}{Pb}, the calculated \twoplus excitation energy is too
high. For \nuc{38}{Ar}, this overestimation has been attributed to
the structure of the state~\cite{Ben03a}. It is indeed very likely
predominantly a broken-pair two-quasiparticle configuration, where
the two occupied magnetic substates in the proton $d_{3/2}$ shell
below the $Z=20$ gap are coupled to $J=2$. A rough estimate for its
excitation energy is provided by two times the proton pairing gap,
which leads to an excitation energy close to the experimental one.
The description of this state requires the breaking of time-reversal
reversal invariance and of axial symmetry in the SCMF, which is
outside of what we can currently handle within our beyond-mean-field
approaches. The excited state in the \nuc{40}{Ca} is of a different
nature. Since, within the shell model, one does not obtain low-lying
even-parity excitations, this state has been famous in the
literature as an early example of shape coexistence. The first
excited state of \nuc{40}{Ca} is a $0^+$ which is the head of an
intruder deformed band. A detailed study of this nucleus with the
full projected GCM has indeed obtained such a band~\cite{Ben03a}.
Although the SLy6 interaction was used in that case, the results
that we find here are very similar.

The next nucleus \nuc{188}{Pb} shows still another kind of behavior.
It is a nucleus with several coexisting minima, which are separated
by tiny barriers only. This isotope has been studied in detail in
Ref.~\cite{Ben04b} with several other neutron-deficient Pb isotopes.
Under the MAP approximation, the ground state is obtained from a
configuration close to sphericity, while the minimum for the first
$2^+$ is strongly oblate. The next example, \nuc{208}{Pb}, is a
heavy doubly magic nucleus. Simple shell model considerations give
already an idea of what should be the dominant component of the
first \twoplus~excitation. It can be obtained by promoting a neutron
from the occupied $i_{13/2}$ shell to the unoccupied $g_{9/2}$
shell, or a proton from the $h_{11/2}$ shell to the $h_{9/2}$ shell.
The single-particle energy differences in the spherical mean-field
configuration are 6.4 and 5.9 MeV, respectively. The MAP energy is 2
MeV higher than the particle-hole energy. Again, like in the case of
\nuc{38}{Ar}, the relevant configurations are broken-pair
two-quasiparticle states outside our configuration space. The last
nucleus in the table, \nuc{240}{Pu}, is highly deformed. Its
character is already seen in the SCMF wave function, which has an
intrinsic mass quadrupole moment of $q=3000$ fm$^2$, of which 1145
fm$^2$ is taken up by the electric quadrupole moment. Assuming that
the wave function corresponds
to a rigid rotor, Eq.~(\ref{eq:rotor_be2}), one
obtains a transition quadrupole moment $\rme = 361$ $e$ fm$^2$~in
agreement with experiment. The MAP approximation does not change
matters; the minimizing $q$ of the $J=0$ and $J=2$ projected
states are very close to that of the SCMF ground state. However,
one sees from the table that the excitation energy of the \twoplus
state is too high by nearly a factor of two. This is another
well-known problem, which has been seen in virtually all
calculations using methods similar to ours: much better agreement
would be obtained using the cranked HFB method to generate a wave
function for the $2^+$ state.

\begin{table}[t!]
\caption{\label{ta:examples}
Results for some selected nuclei.  The reduced transition matrix
element and the spectroscopic quadrupole moment are defined in
Eqs.~\ref{eq:Qred} and~(\ref{eq:Qc}), respectively; the relation
to the $B(E2)$ is given in Eq.~(\ref{eq:B(E2)}). The theoretical
values for \nuc{188}{Pb} from Ref.~\cite{Ben04b} assume that the
ground state is spherical and the excited state is oblate.
}
\begin{tabular}{lccccc}
\hline \noalign{\smallskip}
nucleus & source
        & $E_{ex}$
        & $\lb 0 || \hat{Q}_2 || 2 \rb$
        & $Q_c$
        & \\
        &
        & (MeV)
        & ($e$ fm$^2$)
        & ($e$ fm$^2$)
        & \\
\noalign{\smallskip} \hline \noalign{\smallskip}
\nuc{38}{Ar}  & MAP  & 3.9  &   20.4 &   -22.8   & \\
  & HW-full & 3.7  &   19.9 &   3   & \\
  & HW-6   & 3.8  &   19.1 &   -10   & \\
  & Ref.~\cite{Ben03a}   & 3.6  & 22.8 &  6.9   & \\
 & exp.   & 2.17   &   11.4 &     & \\
\hline
\nuc{40}{Ca}  & MAP   & 5.4  &   23.8 &   34.6   & \\
      & HW-full   & 5.0  &   18.2 &   -6   & \\
     & HW-6   & 5.3  &   16.7 &   1   & \\
  & Ref.~\cite{Ben03a}   & 5.4  & 23.7 &  2.2   & \\
  & experiment    & 3.90  &   9.9 &    & \\
\hline
\nuc{188}{Pb} & MAP   & 0.17  &   192 &   173 & \\
& HW-full & 0.54 &   102 &  170  & \\
& HW-6 & 0.22  &   188 &  180 & \\
& Ref.~\cite{Ben04b} & 0.93  &   71 &  110  & \\
& experiment & 0.72  &    &    & \\
\hline
\nuc{208}{Pb} & MAP   & 7.9  & 99  &   70  & \\
& HW-full & 7.0   & 84    & 1   & \\
& HW-6 & 7.1 &   81 & 6    & \\
& experimental & 4.09  &  55 &    & \\
\hline
\nuc{240}{Pu} & MAP    & 0.076 & 377 & -341  & \\
& HW-full   & 0.076 & 377  & -341   & \\
& HW-6    & 0.076 & 377 & -341   & \\
& Ref.~\cite{Ben04c}   & 0.083 & 377 & -340   & \\
& experimental & 0.043 & 361  &  & \\
\noalign{\smallskip} \hline
\end{tabular}
\end{table}

\subsection{HW and HW-6}

We now examine the effects of configuration mixing on the properties
of the \twoplus state, which are also given in
Table~\ref{ta:examples}. Mixing in the large (``full") configuration
space significantly reduces the energy in two cases, \nuc{40}{Ca}
and \nuc{208}{Pb}, raises the energy in one case, \nuc{188}{Pb}, and
has little or no effect in two cases, \nuc{38}{Ar} and
\nuc{240}{Pu}. The insensitivity for \nuc{240}{Pu} is to be expected
since strongly deformed rotors do not have large shape fluctuations,
see the detailed discussion for this example in Ref.~\cite{Ben04c}.
Including shape fluctuations improves the description of a soft
nucleus as \nuc{188}{Pb}. We see that the full calculation (HW)
produces an excitation energy that approaches the experimental
value. In all the cases where the fluctuations change the energy,
the change goes in the right direction and decreases the theoretical
error.

On the third line of Table~\ref{ta:examples}, we show the effect of
the truncation of the configuration space in the \hbox{HW-6}
approximation. In all but the case of  \nuc{188}{Pb} the energies
are close to the full HW results.  The light Pb isotopes are quite
exceptional, but we saw in the previous section that HW-6 is
reliable enough for a global survey.  The next line in
Table~\ref{ta:examples} shows results from other calculations.  The
reported calculations of \nuc{38}{Ar}, \nuc{40}{Ca},\nuc{188}{Pb}
and \nuc{240}{Pu} were done with the full projected GCM without
approximations, using the same computer codes as here, but with a
slightly different energy functional.  We see that the results are
qualitatively similar to what we found, indicating a mild
sensitivity to the specific energy functional.

We now discuss the quadrupole matrix elements in more detail. The
simplest case is \nuc{240}{Pu}, which, as discussed above, behaves
very much like a rigid rotor.  In the rotor limit, the transition
quadrupole matrix element is proportional to the spectroscopic
quadrupole moment of the \twoplus state. The relation is given in
appendix A, $Q_c / \lb 0|| \hat{Q}_2 || 2 \rb \approx 0.9$. We see
from the table that this is well satisfied for all calculations of
\nuc{240}{Pu}. For the non-deformed nuclei, the spectroscopic
quadrupole moment is small, as would be expected for a spherical
vibrator. In the four cases given in Table~\ref{ta:examples}, the HW
and HW-6 transition matrix elements, although overestimating the
experimental data, are better than a factor two, even in cases where
the dominant component of the $2^+$ appears to be incorrect. This is
probably related to the fact that the quadrupole moment is a bulk
property that is entirely determined by the overall distribution of
the local density, while the energy is sensitive to the detailed
structure and occupation of each single-particle wave function. Note
also that allowing spreading of the wave functions over several
configurations improves the MAP result.

%
%
\section{Global performance}
\label{sect:results}

We carried out the MAP and HW-6 calculations on even-even nuclei with known
binding energies, excluding light nuclei with $N$ or $Z < 8$. This is the
set studied in
Ref.~\cite{Ben06}. Of these, 522 have known \twoplus excitation energies.
These energies range from 39 keV to 6.9
MeV, thus spanning more than 2 orders of magnitude.  The
theoretical numbers span the same range, but as we saw in the last
section there can easily be a factor two error in specific cases.

In view of the results of the previous section, we have excluded
from the full set of nuclei the ones for which one can have
suspicion about our approximation scheme. To identify these
nuclei, we have compared our present HW-6 results with the global
calculation performed earlier where the number of
configurations included in the calculation of the ground state was
not limited. We eliminate all the nuclei for which the
difference between both calculations for the energy of the $0^+$
ground state was larger than 250~keV. The selected set of nuclei
does not include \nuc{188}{Pb}, and similar nuclei which are too
soft to be represented by either a MAP calculation or a small
number of quadrupole configurations. Out of the $522$ nuclei
calculated, there remain after selection \numA.

\subsection{Global results}
\begin{figure}[t!]
\includegraphics{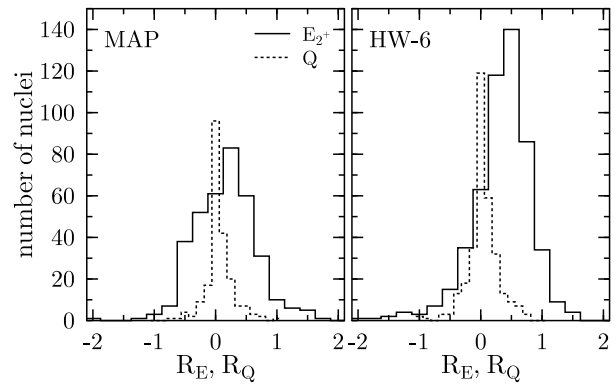}
\caption{Histogram of the logarithmic errors $R_E$ of the
energies of the first excited 2+ states (solid) and the
logarithmic errors $R_Q$ of the reduced matrix elements $\rme$
(dashed) for the \numA even-even nuclei included in our survey
(solid). The two panels show the results of the MAP theory (left)
and the projected GCM in the HW-6 approximation (right).
}
\label{Fi:eq_hist}
\end{figure}

Because the energies span a broad range and the error can be
large, we will quote the aggregated results for the logarithm of
the ratio of the theoretical to experimental energies,
\begin{equation}
R_{E}
=\log(E_{th}/E_{exp}).
\end{equation}
A histogram of this quantity for the
entire set of nuclei is shown in Fig.~\ref{Fi:eq_hist}, displaying
the MAP results in the lefthand panel and the HW-6 results in the
righthand panel.  We see that the results of both methods tend to be too
high, with a fairly broad distribution containing both negative
and positive errors. Quantitative statistical measures of the
distribution are given in Table~\ref{ta:statistics}.

\begin{table}[t!]
\caption{\label{ta:statistics}
Statistics for the performance of the MAP and HW-6 calculations. }
\begin{tabular}{lccccc}
\hline \noalign{\smallskip}
Selection & Number   & observable & theory      & average & dispersion \\
            of nuclei& of nuclei &  &  & $\lb R \rb$ &$ \lb (R-\lb R\rb)^2\rb^{1/2}$  \\
\noalign{\smallskip} \hline \noalign{\smallskip}
all & 359  & $E_{20}$ & MAP  & 0.28 & 0.49 \\
    & 359  & "        & HW-6 & 0.51 & 0.38  \\
    & 212  & $\lb 0 || Q_2 || 2 \rb$ & MAP  &    0.12  &   0.22 \\
    & 212  &  "                      & HW-6 &    0.09  & 0.23 \\
\hline
deformed & 135   & $E_{20}$ &  MAP    & 0.20  & 0.36 \\
         & 135   &  "       &   HW-6  & 0.27  & 0.33 \\
         &  93   & $\lb 0 || Q_2 || 2 \rb$ & MAP   & 0.10 & 0.10 \\
         &  93   & "                       & HW-6  & 0.10 & 0.11 \\
\hline
semi magic & 58 & $E_{20}$ & MAP  & 0.53 & 0.55\\
           & 58 &  "       & HW-6 & 0.58 & 0.31  \\
           & 28 & $\lb 0 || Q_2 || 2 \rb$ & MAP  & 0.37 & 0.24\\
           & 28 &  "                      & HW-6 & 0.35 & 0.23\\
\noalign{\smallskip} \hline
\end{tabular}
\end{table}

\begin{figure}[b!]
\includegraphics{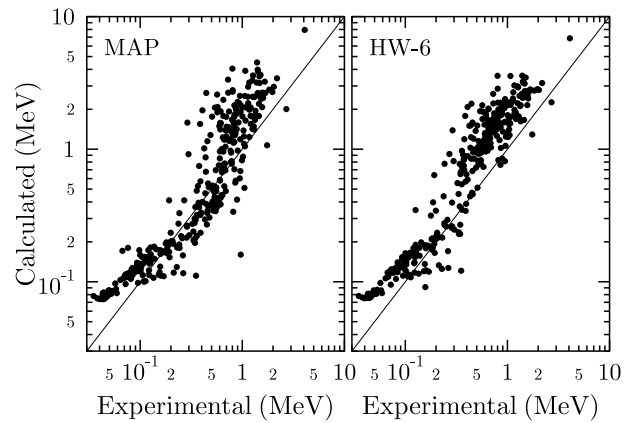}
\caption{Scatter plot comparing the theoretical and experimental \twoplus
excitation energies of the \numA nuclei included in the survey.  The two
panels show the results for the MAP method on the left and the HW-6
approximation on the right.
}
\label{Fi:e_scatter}
\end{figure}

The average MAP error is found to be $\lb R_{E}\rb \approx 0.28 $
but the average of the absolute value of the error is much larger,
 $\lb |R_E|\rb =0.48$, corresponding to an error of the
order of 66 $\%$. The dispersion around the average is also quite
large: $\lb (R_E-\lb R\rb)^2 \rb^{1/2} = 0.49$. With such a
dispersion, an error larger than a factor of two is not unusual.
Specifically, of the \numA nuclei in the data set, 19 $\%$ have a
calculated energy too large by a factor of two and 4 $\%$ are too low
by the same factor. Fig.~\ref{Fi:e_scatter} shows a scatter plot
of the MAP and the experimental energies. One sees that the
energies are overestimated for most nuclei, and in particular for
nuclei with either a low or a high excitation energy of the $2^+$,
where the nuclei are predominantly deformed or magic,
respectively. This is consistent with what we saw in the examples
of the previous section. For excitation energies in the range
200~keV to 1~MeV, there is no obvious trend in error of the MAP
calculation.

As can be seen in Table~\ref{ta:statistics}, the mean error of the
HW-6 calculation is significantly larger than the MAP error, with an
average around $\lb R_{E}\rb \approx 0.51$. The dispersion around
the average is however lower and the average of the absolute value
of the error is only slightly larger than for the MAP results (0.54
compared to 0.48). In Fig.~\ref{Fi:e_scatter} are plotted the HW-6
excitation energies as a function of the experimental data. One can
see that in most cases, the \twoplus excitation is overestimated and
this tendency is much more pronounced than for the MAP results. In
fact, in many cases, the HW-6 $2^+$ energy is larger than the MAP
one. This increase when the configuration mixing correlations are
included means that the correlation energies predicted by our method
are larger in the ground state than in the $2^+$ state. There can be
many origins to this difference in correlation energies. The lack of
triaxial configurations certainly affects more deeply the states
with $J \neq  0$ since for these states the spherical point does not
contribute and the coupling between prolate and oblate
configurations is disfavored. It is also clear that the MAP
procedure is better defined numerically than the HW-6 one. In each
case, we are sure to have determined for both $J=0$ and $J=2$ the
quadrupole moment giving the minimal energy after projection. For
the configuration mixing, the fact that we have excluded the nuclei
for which the $J=0$ energy is too different from our previous global
calculation makes the determination of the $0^+$ energy reliable. We
do not have a similar check for $J=2$ and there are cases where the
number and the spacing of points taken for $J=2$ is not fully
adequate and the energy of this state less accurate.

While the energies are not accurately predicted, the quadrupole
properties come out much better and with rather similar errors for
both the MAP and HW-6 results. It is well known that the intrinsic
quadrupole moments of deformed nuclei are rather insensitive to the
details of the energy functional, and indeed we found that the
quadrupole transition matrix element is much better determined
overall than the energy. The dashed histograms in
Fig.~\ref{Fi:eq_hist} show the logarithmic ratio $R_Q$ of the
reduced quadrupole transition matrix elements $\lb 0 || \hat{Q}_2 ||
2 \rb$ (see appendix \ref{sect:app:Q2}). The average error is only
0.12 for the MAP calculation, corresponding to matrix elements that
are 15\% too large and 0.09 for the HW-6 results (error of 9\%). The
r.m.s. spread is also reduced. For example, in the MAP case, it
takes the value 0.22 corresponding to transition matrix elements
that are -10\% to +46\% of the data.  We will now analyze separately
the different kinds of nuclei. To that aim, we divide the nuclei by
type and examine in more detail the performance for subgroups that
are deformed, doubly-magic, and singly-magic.

\subsubsection{Deformed nuclei}

\begin{figure}[t!]
\includegraphics{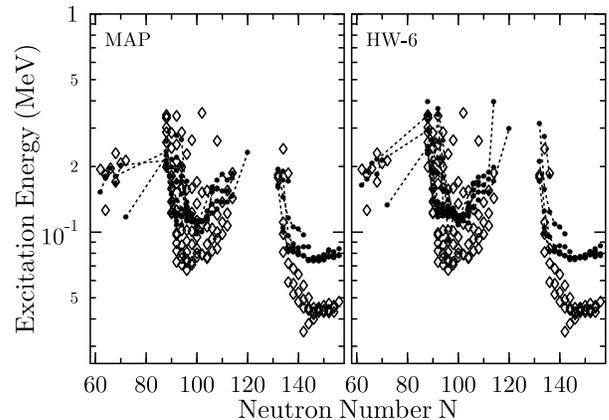}
\caption{\label{fi:e_deformed}
\twoplus excitation energies in deformed nuclei as
a function of neutron number:  MAP in left panel;
HW-6 approximation in right panel. Experimental data are shown as
diamonds.
}
\end{figure}

We first have to define a criterion to select which nuclei should be
considered deformed.  Obviously, there is no rigorous division of
nuclear types, and any division is somewhat arbitrary.  One
possibility is to make a selection on the basis of the intrinsic
quadrupole moment of the MAP ground state, taking into account
overall size effects by using the geometric shape parameter
$\beta_2$ [defined in Eq.~(\ref{eq:beta_2})] to make the selection.
This criterion will catch many light nuclei along with the usual
nuclei in the lanthanide and actinide regions.  One should add the
criterion of rigidity to the selection as well to eliminate the
nuclei that have large fluctuations in shape.  In this sense, what
we are seeking to categorize are nuclei that behave like rigid
rotors. A criterion that makes a nice selection is to demand that
the average deformation $\bar \beta_2$ is large than the r.m.s.
fluctuation about the average, $\bar \beta_2 > \lb (\beta_2
-\bar\beta_2\rb^{1/2}$. These quantities are computed using the full
HW wave functions of Ref.~\cite{Ben06}, and the criterion selects
134 deformed nuclei from our set of 359.  Their energies are plotted
as a function of neutron number in Figs.~\ref{fi:e_deformed} with
the MAP results in the lefthand panel and the HW-6 results in the
righthand panel. The two plots are rather similar.

We see that the predictions are too high for the actinides, while on
the average they are quite reasonable for rare earth nuclei.  The
statistic on the errors for deformed nuclei is summarized in
Table~\ref{ta:statistics}. One can see that the average error is
smaller than for the full set.  The dispersion in the error is the
same for both the MAP and the HW6 approximations, so, the axial
quadrupole correlations do not seem to be the source of the
nucleus-to-nucleus fluctuations of error.

\begin{figure}[t!]
\includegraphics{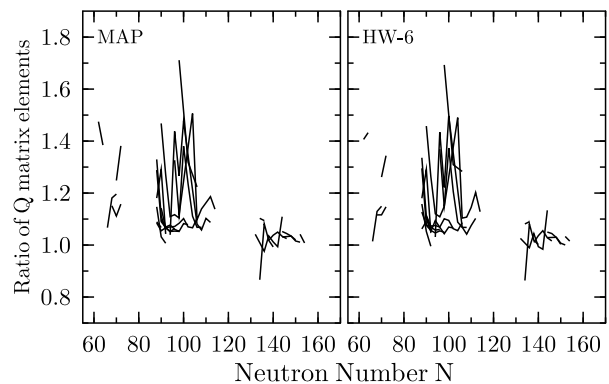}
\caption{Ratio of theoretical and experimental transition quadrupole
matrix elements $\langle 0 || \hat{Q_2} || 2 \rangle$ in deformed
nuclei, as a function of neutron number.
\label{Fi:q_deformedN}
}
\end{figure}

Fig.~\ref{Fi:q_deformedN} shows the ratio of theoretical to
experimental quadrupole transition matrix elements for the deformed
nuclei.  Here the actinide nuclei come out very well.  There is more
fluctuation in the rare earth and the light nuclei that qualify as
strongly deformed but the overall results are quite satisfactory.

\begin{figure*}[t!]
\includegraphics{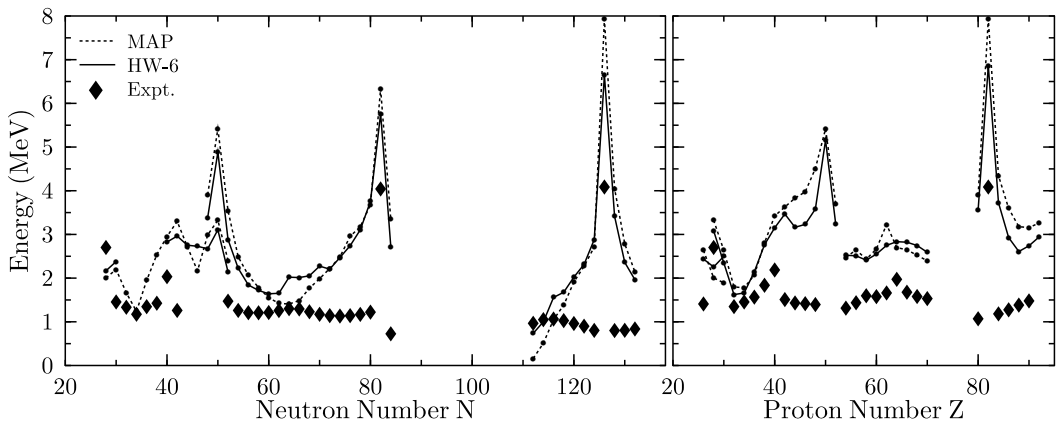}
\caption{\label{Fi:magic}
\twoplus~excitation energies near magic nuclei, as
a function of neutron and proton numbers. MAP and HW6 results
are shown by points connected with lines, while experimental data are
shown by filled diamonds.
}
\end{figure*}

\begin{figure}[t!]
\includegraphics{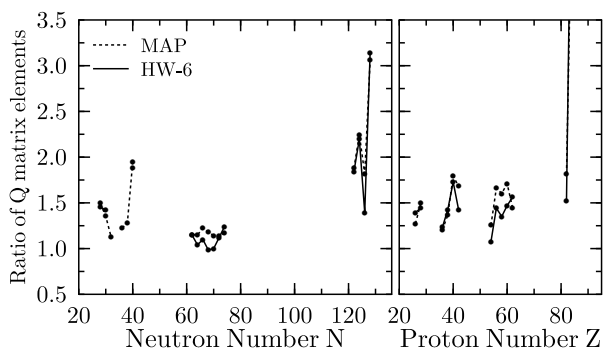}
\caption{\label{Fi:q_magic}
Ratio of theoretical and experimental quadrupole transition
matrix elements near magic nuclei as a function
of neutron and proton numbers. MAP and HW6 results are shown
as points connected by dotted and solid lines, respectively.
}
\end{figure}

\subsubsection{Magic and semi-magic nuclei}

We now turn to doubly- and singly-magic nuclei, which present quite
different problems for the theory.  The comparison between
theoretical and experimental \twoplus excitation energies of six
doubly magic nuclei is shown in Table~\ref{ta:doubly_magic}. The MAP
and HW-6 energies are too high in three cases and too low for the
other three, preventing us from drawing any general conclusions.

\begin{table}[t!]
\caption{\label{ta:doubly_magic}
Excitation energy of the first $2^+$ state in MeV for doubly magic nuclei.}
\begin{tabular}{ccccc}
\hline \hline \noalign{\smallskip}
$N$ & $Z$ & exp. & MAP  & GCM \\ \noalign{\smallskip} \hline \noalign{\smallskip}
20  & 20  & 3.9  & 5.4  & 5.4 \\
28  & 20  & 3.8  & 2.7  & 2.7 \\
28  & 28  & 2.7  & 2.0  & 2.2 \\
82  & 50  & 4.0  & 6.3  & 5.8 \\
126 & 82  & 4.1  & 7.9  & 6.7 \\ \noalign{\smallskip} \hline \hline
\end{tabular}
\end{table}

There are 71 semi-magic nuclei in our compilation, requiring that
either the neutron or the proton number equal to 28, 50, 82 or 126.
Graphs of \twoplus excitation energies are shown in
Figs.~\ref{Fi:magic} as a function of neutron number and proton
number. MAP results are shown by the points connected with solid
lines, HW-6 results with long dashed lines, and experiment by
shorter dashed lines.  For both MAP and HW-6, the excitation energy
has a peak at the doubly-magic nuclides which decreases gradually
going away from that nucleus. In contrast, the experimental peak is
a sharp spike at the doubly-magic nuclides. This deficiency of the
theory is related to the absence of broken-pair two-quasiparticle
excitations that can give a lower non-collective $2^+$ state as we
saw already in the example of \nuc{38}{Ar}. As to be expected from
this discussion, the statistical measures are much poorer for this
class of nuclei. The average calculated energy is 50\% higher than
experiment and the average calculated transition quadrupole moment
is 2.2 times the experimental value.

Fig.~\ref{Fi:q_magic} compares the theoretical and experimental
transition quadrupole moments in semimagic and magic nuclei.
The data is much more meager than for the energies, but one
can see that the theory nearly always is too high.  As discussed
earlier, this is to be expected when the lowest \twoplus is
not collective.

\subsection{Discussion}

In view of their restricted form and of the way they have been
fitted to selected experimental data, the current energy functionals
are certainly too limited and, as discussed in \cite{Ben06,Ben06b},
present deficiencies which are at the origin of some of the
discrepancies between our calculations and experiment. However, the
present analysis clearly points also to deficiencies of the
variational space that is used, that affect more excitation energies
than quadrupole moments.

One can expect that our configuration space spanned by axial
quadrupole SCMF wave functions covers the correlations that dominate
the description of the $0^+$ ground states of even-even nuclei. On
the contrary, there are several competing possibilities to construct
a low-lying $2^+$ state, some of them being completely absent from
our description:
\begin{enumerate}
\item
a broken-pair two-quasiparticle excitation within a partly-filled $j$ shell
with an excitation energy of about two times the pairing gap
(for example near-magic nuclei like \nuc{38}{Ar})
\item
a broken pair two-quasiparticle excitation involving two different
$j$ shells, one occupied, the other unoccupied,
with an excitation energy of about the gap between the $j$
shells involved (for example doubly-magic nuclei as \nuc{208}{Pb})
\item
a collective vibrational state
\item
a collective rotational state (for example for well-deformed
rare-earth and actinide nuclei)
\end{enumerate}
The states corresponding to such pure configurations should of
course be mixed in actual nuclei. The projected GCM of axially
deformed SCMF states that correspond to HFB vacuua, as it is used
here, cover only the latter two of these configurations. In nuclei
where the lowest $2^+$ state is dominated by broken-pair
two-quasiparticle states, the $2^+$ state that our method enables us
to describe corresponds to a higher-lying collective $2^+$ state. On
the other hand, the number of nuclei for which the lowest $2^+$
state is indeed dominated by a broken-pair two-quasiparticle state
can be expected to be small, and restricted to the immediate
vicinity of doubly-magic nuclei. In nuclei where the first $2^+$
is expected to be collective, either vibrational, or rotational,the
excitation energies are also on the average too high. This result
confirms on a large scale previous studies performed with similar
methods for smaller sets of nuclei
\cite{Ben03a,Ben04b,Ben04c,Ben05b,ro02,Rod02a,Egi04a}

An obviously missing degree of freedom is triaxiality, as our
configuration mixing contains only two out of the five degrees of
freedom of the quadrupole tensor. On the one hand, it is
well-known that many transitional nuclei are $\gamma$ soft. A
recent global study based on a semi-microscopic method has even
indicated that the potential energy surface of many transitional
nuclei might have a triaxial minimum~\cite{Moe06a}, although with
an energy gain that remains very small. It has also been
shown~\cite{Har82a} that in some cases triaxial quadrupole
configurations can be more favorable after angular-momentum
projection than axial configurations. The effect of triaxiality on
excitation energies has also been studied with the help of an
effective five-dimensional Bohr Hamiltonian derived from
mean-field calculations using the Gogny force~\cite{Gir78a,li99}
or a Skyrme interaction \cite{Pro04a}. A similar overestimation
of the lowest $2^+$ energy as in the present study has been found.
Therefore, if the effects of triaxiality are certainly
non-negligible on total energies, it is unclear whether they
will improve excitation energies.

\begin{table}[t!]
\caption{\label{ta:cranking}
Excitation energy of the first $2^+$ state in MeV for Zn isotopes.}
\begin{tabular}{lcccc}
\hline \hline \noalign{\smallskip}
 $N$   & Expt.       & MAP        & HW-6      & cranked SCMF  \\
\noalign{\smallskip} \hline \noalign{\smallskip}
 30    & 0.89        & 1.85       & 1.89      & 1.35      \\
 32    & 0.56        & 2.36       & 1.94      & 0.90      \\
 34    & 0.61        & 1.56       & 1.81      & 0.41      \\
 36    & 0.60        & 1.09       & 1.71      & 0.37      \\
 38    & 0.73        & 1.88       & 1.95      & 0.41      \\
 40    & 1.50        & 2.64       & 2.33      & 0.80      \\
\noalign{\smallskip} \hline \hline
\end{tabular}
\end{table}

The effect of triaxial quadrupole deformations on excitation
energies is not obvious. The method that we use has however clearly
an artifact which favors the $0^+$ energy with respect to the $2^+$:
the mean-field is optimized to describe the ground state and not
excited states. An obvious improvement would be to perform an exact
variation after projection (VAP), performed separately for $J=0$ and
$J=2$. Variation after projection on angular momentum starting from
effective interactions and with a full model space does not seem
however within computational possibilities in a near future. A more
modest approach would be to use the self-consistent cranking method
to optimize
separately the intrinsic configurations describing different spin
states. This can be done by introducing in the mean-field equations
a constraint on the projection of the angular momentum. It has been
shown that the self-consistent cranking method is an approximation
of a variation after projection on angular momentum for well
deformed nuclei~\cite{RS80}. Such an approach will necessarily
improve the description of the $2^+$ states and decrease their
excitation energy.

Even for nuclei which are not deformed, the effect of a cranking
constraint, which breaks time-reversal invariance can only go in the
right direction, although it is not clear \emph{a priori} that the
introduction of two-quasiparticle excitations would not be even more
important. To give some
insights into the effect of a cranking constraint for isotopes which
are not deformed, we show in Table~\ref{ta:cranking} the results of
MAP, HW-6 and cranked SCMF calculations together with the
experimental data for some neutron-rich Zn isotopes. The cranked
SCMF calculations were done with the method described in Ref.\
\cite{95e} and with the same effective interaction as for the other
calculations of this study. Both MAP and HW-6 results strongly
overestimate the experimental energies, the MAP being even rather
irregular in its predictions. The cranking results are given in the
last column. The $2^+$ excitation energy in this case is the
difference between the mean-field ground state energy and that of
the state obtained with a cranking constraint $J_z=2$. One sees that
the energies are significantly lower compared to the calculations
where time reversal invariance is imposed. The numbers that are
obtained can even be lower than the experimental data. One cannot go
too far in the interpretation of these results which do not include
any projections. However, they show that an optimization of the
$2^+$ wave function with a cranking constraint might have a
significant effect going in the right direction for all nuclei.

\begin{figure}[t!]
\includegraphics{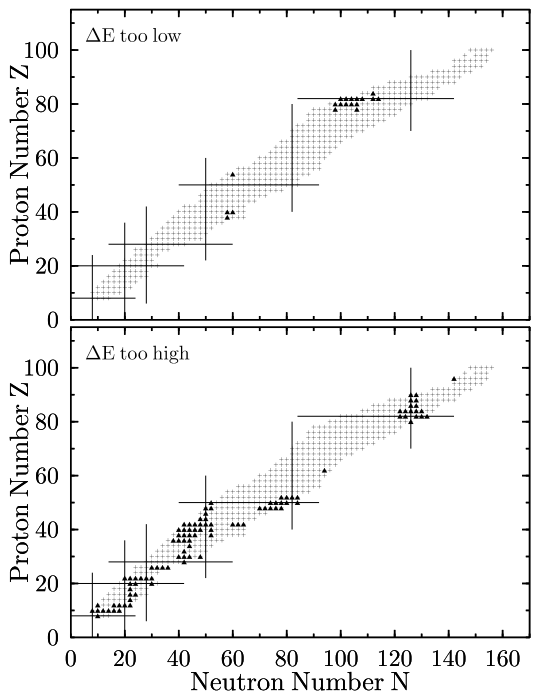}
\caption{\label{fi:e_worst}
Chart of nuclides showing the even-even nuclei for which
the excitation energy of the first \twoplus state is known.
Nuclei for which the MAP calculation is in error by more than a
factor of two are shown by solid triangles. Upper panel: theory too low;
lower panel: theory too high.
}
\end{figure}

\begin{figure}
\includegraphics[width=7cm]{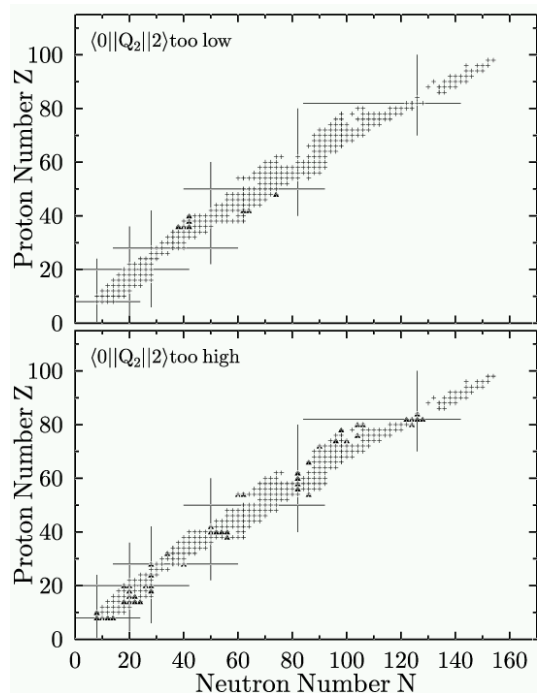}
\caption{\label{fi:be2_worst}
Chart of nuclides showing the even-even nuclei for which
the transition quadrupole moment to the first \twoplus state is known.
Nuclei for which the MAP calculation is in error by more than a
factor of two are shown by solid triangles. Upper panel: theory too low;
lower panel: theory too high.
}
\end{figure}

It remains to verify what will happen when exact projection and
configuration mixing will be performed, but the perspective of
a global qualitative improvement of the present results seems
reasonable.
The generalization of the method used here for use with cranked SCMF
states requires a similar, even greater, effort as the generalization
to triaxial shapes: the cranking constraint induces non-axial intrinsic
currents, even when the overall density distribution remains axial.
The broken-pair two-quasiparticle states discussed above
require a similar generalization of the spatial symmetries in the
projected GCM as the cranked HFB vacuum.

%
%
\section{Summary and outlook}
\label{sect:summary}

This is the first systematic survey of the $2^+$ excitations of
even-even nuclei in the framework of a mean-field based method
including symmetry restoration and starting from an effective energy
functional. The effort necessary for this task is significantly
larger than the one required for our earlier study on the ground
states of these nuclei \cite{Ben06}, both for the representation of
the collective $2^+$ states, and to obtain a sufficient precision
for the matrix elements of the quadrupole operator. For several
nuclei we could not reach an acceptable precision, so that the
subsequent analysis was performed on a reduced set of data.

Qualitatively the excitation energies and $B(E2)$ values track
the data for the great majority of the \numA nuclei studied.
However, predicted energies and $B(E2)$ values are systematically too
high, and there are a number of cases where key ingredients are
clearly missing. The worst cases, where the observable is more
than a factor of two in error, are marked on the charts of
nuclides in Figs.~\ref{fi:e_worst} and~\ref{fi:be2_worst}. One can
discern some patterns that point to deficiencies in the energy
functional and in the GCM methodology that both may be correctable.

Many of the outlying points in Fig.~\ref{fi:e_worst} are cases
where the theory predicts a nearly spherical nucleus while the
data show it to be deformed, or vice versa. An example is
\nuc{80}{Zr}, predicted spherical but obviously deformed in view
of the very low excitation energy of its first $2^+$. The shell
effect predicted by our effective interaction for \nuc{80}{Zr} is
clearly too large, as already analyzed in our study of
ground-state correlations~\cite{Ben06}. All conclusions of
Ref.~\cite{Ben06} about necessary future work on the effective
interactions and the model space also apply here, see
also~\cite{Ben06b}.

Our calculation reproduces rather nicely the quadrupole transition
matrix elements between the first $2^+$ and the ground state: the
average error that we obtain is around 25$\%$. The situation is
less satisfactory for the excitation energies of the first $2^+$
states, which is nearly always overestimated. This seems to be a
general problem that has been noticed before in many calculations
using Skyrme and Gogny interactions. As argued above, we relate
this deficiency mainly to the current restrictions of the
variational space that we use. To overcome this limitation, the
extension of the variational space to include triaxial states, and
cranked SCMF states is highly desirable. Work in that direction is
underway. The enormous increase in computational time, however,
will not permit its large-scale application right away.
%
%
\section*{Acknowledgments}
We thank A. Bulgac, H. Goutte, W. Nazarewicz, and P.-G. Reinhard
for discussions.
Financial support was provided by the U.S.\ Dept.\ of Energy under
Grants DE-FG-02-91ER40608 and DE-FG02-00ER41132 (Institute for Nuclear
Theory), and the Belgian Science Policy Office under contract PAI P5-07.
Part of the work by M.~B.\ was performed within the framework of
the Espace de Structure Nucl{\'e}aire Th{\'e}orique (ESNT).
The computations were performed at the National Energy Research
Scientific Computing Center, supported by the U.S.\ Dept.\ of
Energy under Contract No.\ DE-AC03-76SF00098.
%
%
\appendix

\section{Matrix elements of the quadrupole operator}
\label{sect:app:Q2}

For reference, we quote the definitions of the quadrupole matrix
elements and simplified versions of the formulas from Ref.~\cite{Ben03a}
for calculating them. For the sake of simple notation, we will give
all expressions, where applicable, for matrix elements between two
different SCMF configurations, $|q \rb$ and $|q' \rb$. The
generalization to GCM states with their weighted summation is
straightforward as it does not affect the angular momentum algebra.

The electric quadrupole operator is defined as
\begin{equation}
\label{eq:Qdef}
\hat{Q}_{2\mu}
= e \sum_{p} r^2_i Y_{2\mu}(\hat r_i)
\end{equation}
We start with Eq.~(A7) of Ref.~\cite{Ben03a} for the reduced
matrix element of the quadrupole operator between two projected
axial states
\begin{eqnarray}
\label{eq:Qred}
\lb J q || \hat{Q}_2 || J' q' \rb
& = & \frac{\sqrt{2 J+1}(2J'+1)}
           {\mathcal{N}_{Jq} \mathcal{N}_{J'q'}}
      \sum_{\mu=-2}^{+2}
      \lb J'0 2 \mu | J \mu\rb
      \nn \\
&   & \times
      \int_0^1 \! d \cos (\beta) \, d^J_{0 \mu}(\beta) \;
      \lb q | \hat{R}_\beta \hat{Q}_{2\mu} | q' \rb
      \nn \\
\end{eqnarray}
where $J$ and $J'$ are assumed to be integer and even.
$\hat{R}_\beta$ is the rotation operator, $d^J_{\mu\kappa}(\beta)$
is the Wigner $d$-function, and $\mathcal{N}_{Jq}$ is the normalization
of the $J$-projected SCMF state, Eq.~(\ref{eq:norm:proj}).
The reduced matrix element on the left-hand side is defined
as~\cite{Var88a}
\be
\lb J M q | \hat{Q}_{2\mu} | J' M' q' \rb
= \frac{\lb J' M' 2 \mu | J M\rb}{\sqrt{2 J + 1}}
  \lb J q || \hat{Q}_2 || J' q' \rb
.
\end{equation}
Equation (\ref{eq:Qred}) can be even simplified further using the
symmetries of the Wigner functions and the quadrupole operators
\begin{widetext}
\begin{eqnarray}
\label{eq:Qred:2}
\lb J q || \hat{Q}_2 || J' q' \rb
& = & \frac{\sqrt{2 J+1}(2J'+1)}
           {\mathcal{N}_{Jq} \mathcal{N}_{J'q'}}
      \bigg(   \lb J'0 2 0 | J 0 \rb
               \int_0^1 \! d \cos (\beta) \; d^J_{00}(\beta) \;
               \lb q | \hat{R}_\beta \hat{Q}_{20} | q' \rb
     \nn \\
&  & \qquad \qquad \qquad
             + 2 \sum_{\mu=1}^{2}
               \lb J'0 2 \mu | J \mu \rb
               \int_0^1 \! d \cos (\beta) \; d^J_{0\mu}(\beta) \;
               \Re \{ \lb q | \hat{R}_\beta \hat{Q}_{2\mu} | q' \rb \}
       \bigg)
,
\end{eqnarray}
\end{widetext}
which serves as the starting point for the GOA set-up in
section \ref{sect:Q2:mat}.

To compute the matrix element for the $2^+ \to 0^+$
transition, one evaluates the above formula with $J=0$ and $J'=2$.
Only $\mu=0$ contributes in this case and the result is
\begin{equation}
\label{eq:Q2red:trans}
\lb 0 q || \hat{Q}_2 || 2 q' \rb
= \frac{\sqrt{5}}{\mathcal{N}_{2q} \mathcal{N}_{0q'}}
  \int^1_{0} \! d \cos(\beta) \;
  \langle q | \hat{R}_\beta \hat{Q}_{20} | q' \rangle
.
\end{equation}
The $B(E2)$ for the transition is related to the reduced matrix
element by \cite{Var88a}
\begin{equation}
\label{eq:B(E2)}
B(E2,2^+ \to 0^+)
= \frac{1}{5} \; \langle 0 q || \hat{Q}_2 || 2 q' \rangle^2
.
\end{equation}
The other matrix element of interest is the spectroscopic quadrupole
moment of the $J=2$ excited state, defined as
\begin{equation}
Q_c
= \sqrt{\frac{16 \pi}{5}} \langle 2 2 q | \hat{Q}_{20} | 2 2 q' \rangle
.
\end{equation}
In this case the sum over $\mu$ cannot be avoided.  The final result
is
\begin{equation}
\label{eq:Qc}
Q_c
= -\frac{5}{\mathcal{N}_{2}^2}
  \left( - \frac{2}{7} M_0
         + \frac{2}{7} M_1
         + \frac{4}{7} M_2
  \right)
\end{equation}
where
\begin{equation}
\label{eq:Mmu}
M_\mu
= \sqrt{\frac{16 \pi}{5}}
  \int^1_{0} \! d\cos(\beta) \;
  d^2_{0 \mu }(\beta) \;
  \langle q | \hat{R}_\beta \hat{Q}_{2\mu} | q' \rangle
.
\end{equation}
The rotor model provides a convenient reference for estimating quadrupole
matrix elements.
In terms of the intrinsic quadrupole moment of the configuration,
$\lb q | \hat{Q}_{20}| q\rb$, the relations are
\be
\label{eq:Qc_rotor}
Q_{c,\text{rotor}}
= -{2\over 7} \sqrt{\frac{16 \pi}{5}} \lb q | \hat{Q}_{20}| q \rb
\ee
and
\be
\lb 0 q || \hat{Q}_2 || 2 q \rb_{\text{rotor}}
= \lb q | \hat{Q}_{20}| q \rb
\label{eq:rotor_be2}
\ee
Finally, we specify the deformation of a configuration by the mass
quadrupole moment with the spectroscopic normalization.  The relation is
\be
\label{eq:qdef}
q
= \frac{1}{2} \,
  \lb q |  \sum_{n,p} r_i^2 \, [ 3 \cos^2 (\theta) - 1 ] \, |q \rb
.
\ee
We also use the dimensionless deformation parameter $\beta_2$ defined
by the equation
\be
\label{eq:beta_2}
\beta_2
= \frac{\sqrt{5 \pi}}{3} \frac{q}{A R_0^2},
\ee
using the liquid drop radius constant \mbox{$R_0 = 1.2 \; A^{1/3}$} fm.
%
%
\section{Selection of configurations in HW-6}
\label{sect:app:HW6}

In this appendix we describe in more detail how the configuration
set was chosen for the configuration mixing.  The following rules
were applied to select configurations for each nucleus and for
$J=0$ and $J=2$. The rules are:
\begin{itemize}
\item
Start with the set of 15-20 constrained configurations
that were used in our previous study ~\cite{Ben06}.
\item
Divide the set into prolate and oblate configurations.  For
both sets and each angular momentum value, find the configurations
$|q_{\text{min}}\rangle$ that have the minimum energy after
particle-number and angular-momentum projection.
\item
In each set, select the projected configurations on
each side of the minima that have an overlap close to, but larger
than 0.5 with $|q_{\text{min}}\rangle$. This leaves both sets with
up to three configurations.
\item
Join the prolate and oblate sets, taking out
the oblate configuration with the lowest deformation if its
overlap with the least deformed prolate configuration is greater
than 0.9.
\item
Add to the set of $J=0$ configurations all the $J=2$ configurations
that do not overlap a $J=0$ configuration by more than 0.9. Likewise
add $J=0$ configurations to the $J=2$ set.
\end{itemize}
Most resulting sets include 5 to 6 configurations, although some
could be larger or smaller. For example, for nuclei in the rare-earth
and actinide regions that present a deep and narrow prolate minimum in the
total energy surface, the oblate configurations
are too high in energy to play a role and 3 configurations are
sufficient.  For some other nuclei, several points are needed to
connect the prolate and oblate sets, making the configuration set
larger than 6. In a few cases, the selected sets lead to
instabilities in the solution of the HW equations, related to too
small eigenvalues of the norm kernel. These cases had to be
treated by hand to select the configurations.
%
%


\begin{thebibliography}{99}

\bibitem{RMP}
  M. Bender, P.-H. Heenen, and P.-G. Reinhard,
  Rev. Mod. Phys. \textbf{75}, 121 (2003).

\bibitem{doba}
  M. V. Stoitsov, J. Dobaczewski, W. Nazarewicz, S. Pittel, D.J. Dean,
  Phys. Rev. C \textbf{68}, 054312 (2003);
  M. V. Stoitsov, J. Dobaczewski, W. Nazarewicz, P. Borycki,
  Int. Jour. Mass Spectr. \textbf{251}, 243 (2006).

\bibitem{Ben05a}
  M. Bender, G. F. Bertsch, and P.-H. Heenen,
  Phys. Rev. Lett. \textbf{94}, 102503 (2005).

\bibitem{Ben06}
  M. Bender, G. F. Bertsch, and P.-H. Heenen,
  Phys. Rev. C \textbf{73}, 034322 (2006).

\bibitem{O-16}
  M. Bender and P.-H. Heenen,
  Nucl. Phys. \textbf{A713}, 390 (2003).

\bibitem{Ben03a}
  M. Bender, H. Flocard and P.-H. Heenen,
  Phys. Rev. C \textbf{68}, 044321 (2003).

\bibitem{Ben04b}
  M. Bender, P. Bonche, T. Duguet, and P.-H. Heenen,
  Phys. Rev. C \textbf{69}, 064303 (2004).

\bibitem{Ben04c}
  M. Bender, P.-H. Heenen, P. Bonche,
  Phys. Rev. C \textbf{70}, 054304 (2004).

\bibitem{Ben05b}
  M. Bender and P.-H. Heenen,
  Proceedings of ENAM'04,
  C. Gross, W. Nazarewicz, and K. Rykaczewski [edts.],
  Eur. Phys. J. A \textbf{25} s01, 519 (2005).

\bibitem{ro02}
  R. Rodriguez-Guzman, J. L. Egido, and L. M. Robledo,
  Phys. Rev. C \textbf{65}, 024304 (2002).

\bibitem{Rod02a}
  R. Rodriguez-Guzman, J. L. Egido, and L. M. Robledo,
  Nucl. Phys. \textbf{A709}, 201 (2002).

\bibitem{Egi04a}
  J. L. Egido and L.M. Robledo, in
  \emph{Extended Density Functionals in Nuclear Physics},
  G. A. Lalazissis, P. Ring, D. Vretenar [edts.],
  Lecture Notes in Physics No. 641 (Springer, Berlin, 2004),
  p. 269.

\bibitem{Nik06a}
  T. Nik{\v s}i{\'c}, D. Vretenar, and P. Ring,
  Phys. Rev. C \textbf{73}, 034308 (2006).

\bibitem{Nik06b}
T. Nik{\v s}i{\'c}, \etal  Phys. Rev. C, in print;
  preprint nucl-th/0611022.

\bibitem{re78}
  P.-G. Reinhard,
  Z. Phys. A \textbf{285} 93 (1987).

\bibitem{ha02}
  K. Hagino, P.-G. Reinhard, and G. F. Bertsch,
  Phys. Rev. C \textbf{65} 064320 (2002).

\bibitem{Bon85a}
  P. Bonche, H. Flocard, P.-H. Heenen, S. J. Krieger, and M. S. Weiss,
  Nucl. Phys. \textbf{A443}, 39 (1985).

\bibitem{Bon05a}
  P. Bonche, H. Flocard, and P.-H. Heenen,
  Computer Phys. Comm. \textbf{171}, 49 (2005).

\bibitem{Cha98}
  E. Chabanat, P. Bonche, P. Haensel, J. Meyer, and R. Schaeffer,
  Nucl. Phys. \textbf{A635}, 231    (1998),
  Nucl. Phys. \textbf{A643}, 441(E) (1998).

\bibitem{Rig99}
  C. Rigollet, P. Bonche, H. Flocard, and P.-H. Heenen,
  Phys. Rev. C \textbf{59}, 3120 (1999).

\bibitem{Gal94a}
  B. Gall,  P. Bonche, J. Dobaczewski, H. Flocard, and P.-H. Heenen,
  Z. Phys. \textbf{A348}, 183 (1994).

\bibitem{promesse}
  M. Bender, P. Bonche, and P.-H. Heenen,
  unpublished.

\bibitem{Ben04a}
  M. Bender, G. F. Bertsch, and P.-H. Heenen,
  Phys. Rev. C \textbf{69}, 034340 (2004).

\bibitem{ha03}
  K. Hagino, G. F. Bertsch, and P.-G. Reinhard,
  Phys. Rev. C \textbf{68}, 024306 (2003).

\bibitem{Ben06b}
  M. Bender, P. Bonche, P.-H. Heenen,
  Phys. Rev. C \textbf{74}, 024312 (2006).

\bibitem{ra02}
  S. Raman, C. W. Nestor, Jr., and P. Tikkanen,
  At. Data Nucl. Data Tables \textbf{78}, 1 (2001).

\bibitem{Moe06a}
  P. M{\"o}ller, R. Bengtsson, B. G. Carlsson, P. Olivius, and T. Ichikawa,
  Phys. Rev. Lett. \textbf{97}, 162502 (2006).

\bibitem{Har82a}
  K. Hara, A. Hayashi, and P. Ring,
  Nucl. Phys. \textbf{A385}, 14 (1982).

\bibitem{Gir78a}
  M. Girod and K. Kumar, B. Grammaticos, and P. Aguer,
  Phys. Rev. Lett. \textbf{41}, 1765 (1978).

\bibitem{li99}
  J. Libert, M. Girod, and J.-P. Delaroche,
  Phys. Rev. C \textbf{60}, 054301 (1999).

\bibitem{Pro04a}
  L. Pr{\'o}chniak, P. Quentin, D. Samsoen and J. Libert,
  Nucl. Phys. \textbf{A730}, 59 (2004).

\bibitem{RS80}
  P. Ring and P. Schuck,
  \emph{The Nuclear Many Body Problem},
  (Springer, Berlin, 1980).

\bibitem{95e}
  J. Terasaki, P.-H. Heenen, P. Bonche, J. Dobaczewski, and H. Flocard,
  Nucl. Phys. \textbf{A593}, 1 (1995).

\bibitem{Var88a}
  D. A. Varshalovich, A. N. Moskalev, V. K. Khersonskii,
  \emph{Quantum Theory of angular momentum},
  World Scientific, Singapore, 1988.






\end{thebibliography}
\end{document}